\documentclass[acmtog]{acmart}

\usepackage{algorithm}
\usepackage{algpseudocode} 
\usepackage{multirow}
\usepackage{amsmath}
\usepackage{colortbl}

\AtBeginDocument{%
  }

\setcopyright{acmlicensed}
\copyrightyear{2025}
\acmYear{2025}
\acmDOI{XXXXXXX.XXXXXXX}
\acmConference[SIGGRAPH Asia 2025]{SIGGRAPH Asia 2025}{December 15–18, 2025}{Hong Kong, Hong Kong}%

\acmISBN{978-1-4503-XXXX-X/2018/06}

\acmSubmissionID{1846}

\begin{document}

\title{Rigidity-Aware 3D Gaussian Deformation from a Single Image}

\author{Jinhyeok Kim}
\email{jinhyeok@unist.ac.kr}
\affiliation{%
  \institution{UNIST}
  \country{Republic of Korea}
}

\author{Jaehun Bang}
\email{devappendcbangj@unist.ac.kr}
\affiliation{%
  \institution{UNIST}
  \country{Republic of Korea}
}

\author{Seunghyun Seo}
\email{gogogo0312@unist.ac.kr}

\affiliation{%
  \institution{UNIST}
  \country{Republic of Korea}
}

\author{Kyungdon Joo\textsuperscript{†}}
\email{kdjoo369@gmail.com}
\thanks{†~Corresponding author}

\affiliation{%
  \institution{UNIST}
  \country{Republic of Korea}
}

\renewcommand{\shortauthors}{Trovato et al.}

\newcommand{\modified}[1]{\textcolor{black}{#1}}
\newcommand{\best}{\cellcolor{red}}
\newcommand{\sbest}{\cellcolor{orange}}
\newcommand{\tbest}{\cellcolor{yellow}}

\begin{abstract}
    Reconstructing object deformation from a single image remains a significant challenge in computer vision and graphics. 
    Existing methods typically rely on multi-view video to recover deformation, limiting their applicability under constrained scenarios. 
    To address this, we propose \texttt{DeformSplat}, a novel framework that effectively guides 3D Gaussian deformation from only a single image. 
    Our method introduces two main technical contributions. 
    First, we present Gaussian-to-Pixel Matching which bridges the domain gap between 3D Gaussian representations and 2D pixel observations. This enables robust deformation guidance from sparse visual cues. 
    Second, we propose Rigid Part Segmentation consisting of initialization and refinement. 
    This segmentation explicitly identifies rigid regions, crucial for maintaining geometric coherence during deformation. By combining these two techniques, our approach can reconstruct consistent deformations from a single image. 
    Extensive experiments demonstrate that our approach significantly outperforms existing methods and naturally extends to various applications,
    such as frame interpolation and interactive object manipulation.
    Project page :
    \textcolor{magenta}
    {\texttt{\href{https://vision3d-lab.github.io/deformsplat}{https://vision3d-lab.github.io/deformsplat}}}
\end{abstract}


\begin{teaserfigure}
    \centering
    \includegraphics[width=0.95\linewidth]{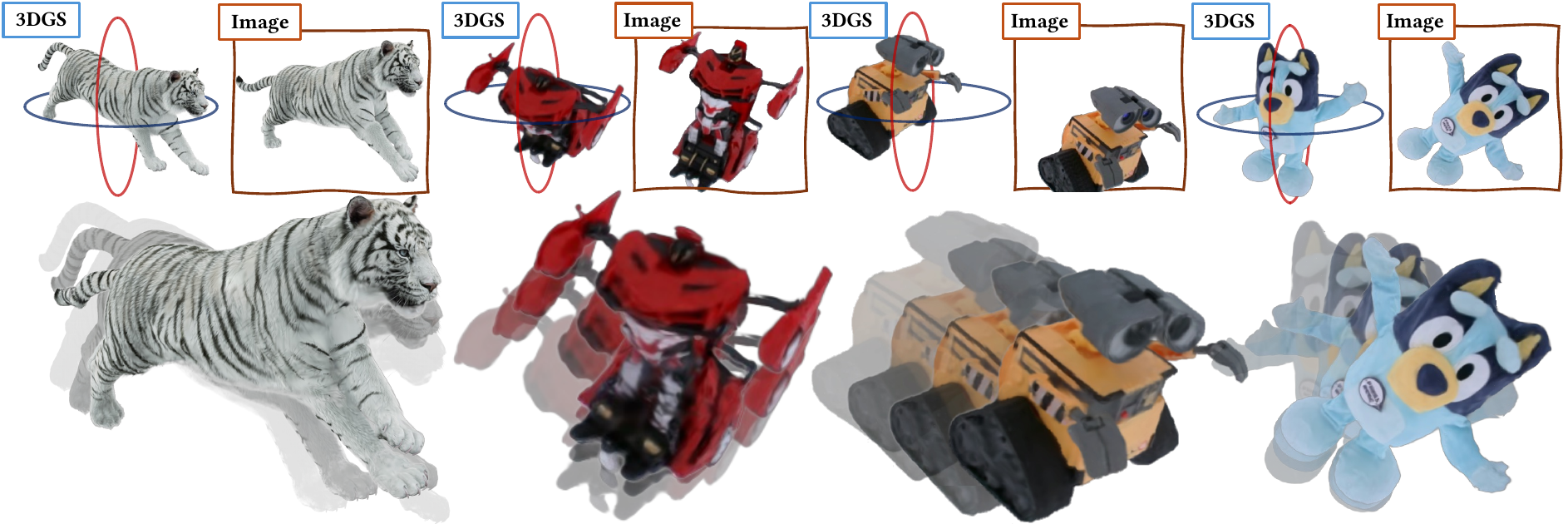}
    \caption{\textit{Overview of our task.} Given a single target image and an initial 3D Gaussian, \texttt{DeformSplat} deforms the Gaussian to match the target image while preserving geometry. The motion is represented by varying the transparency of the images over time.}
    \Description{Enjoying the baseball game from the third-base seats. Ichiro Suzuki preparing to bat.}
  \label{fig:teaser}
\end{teaserfigure}


\maketitle

\section{Introduction}

Reconstructing object deformation from visual data
is essential for creating realistic and immersive content in various media fields, such as virtual reality (VR), film, and gaming~\cite{shuai2022novel}.
%
As part of this effort, recent works have explored photorealistic rendering of deformable objects, aiming to better capture their appearance and motion over time~\cite{wu20244dgs,lu2024gags}.
%
Although these methods have demonstrated strong capabilities for dynamic scene reconstruction, they often rely on multi-view or temporally continuous video data.
Such data can be difficult to capture in real-world settings, which limits their practical applicability.
This motivates the development of methods that reconstruct deformation from minimal visual inputs.

To achieve high-quality and efficient reconstruction of deformable objects, recent research has focused on scene representations that support both photorealism and editability.
Among these, 3D Gaussian Splatting (3DGS)~\cite{kerbl20233dgs} has gained attention for its high-quality rendering and fast inference.
Unlike implicit neural representations, such as Neural Radiance Fields (NeRF) \cite{mildenhall2021nerf}, 3DGS explicitly models geometry, making it both interpretable and easy to manipulate.
Motivated by these advantages, several subsequent works  have explored Gaussian-based scene editing. 
For example, several approaches leveraging diffusion models enable intuitive editing guided by text prompts \cite{wu2024gaussctrl} or reference images \cite{mei2024regs}. 
%
While these approaches enable intuitive edits, diffusion models often produce inconsistent results, as ambiguous text prompts lead to varied interpretations and limited geometric control.
Another recent method, GESI~\cite{luo2024gesi}, aims to address detailed geometric editing directly using a single reference image. 
%
However, it encounters difficulties in preserving intricate geometry under long-range deformations, often altering the original structure significantly.
%
These limitations motivate our key research question: \emph{Can 3D Gaussians be deformed from a single image while preserving the original geometry?}

In this work, we aim to deform a pre-reconstructed 3D Gaussian representation using only a single target image depicting a deformation, as illustrated in Fig.~\ref{fig:teaser}.
%
%
Our setting is challenging, as we aim to deform 3D Gaussians using only a single RGB image, unlike conventional methods
that rely on richer inputs such as multiple views or video sequences.
The absence of depth and camera pose information further complicates the deformation process in our case.
%
This constrained setting gives rise to two key challenges.
%
The first challenge is determining how and in which direction the Gaussians should deform when only a single viewpoint is available.
%
This is difficult because a single image provides only partial observations of the 3D structure, making it hard to infer meaningful deformation cues.
To extract meaningful deformation cues under this constraint, it is essential to establish reliable correspondences between the 2D image and the 3D Gaussians.
%
The second challenge is to prevent overfitting to the single input image, which can result in unwanted geometric distortions due to the lack of depth or multi-view constraints.
%
%
Thus, preserving original geometry is crucial, especially in rigid regions that should remain unchanged during deformation.

To address these challenges, we propose a novel framework called \texttt{DeformSplat}, consisting of two main components: Gaussian-to-Pixel Matching and Rigid Part Segmentation. 
Gaussian-to-Pixel Matching aims to guide the deformation by linking visually similar regions between the 3D Gaussians and the target image.
%
Specifically, we render multi-view images from the 3D Gaussians and compute pixel-wise correspondences between each rendered image and the target image using an image matcher.
%
Based on the pixel-to-pixel matching, we select the viewpoint with the largest visual overlap, and translate its correspondences into Gaussian-to-Pixel mappings.
This step provides an essential basis for deformation, enabling the Gaussians to reflect the geometry depicted in the target image.

To further ensure geometric consistency, we propose Rigid Part Segmentation that explicitly identifies rigid regions within the Gaussian representation.
To achieve this, our method first initializes rigid groups based on Gaussian-to-Pixel correspondences and then iteratively refines these groups during optimization.
%
The segmentation is used in a rigidity-aware optimization that regularizes rigid and non-rigid regions differently to preserve geometry.
Consequently, we achieve superior performance than previous SOTA and generalize to applications such as frame interpolation and interactive object manipulation.

Our contributions can be summarized as follows:
\begin{itemize}
\item We propose \texttt{DeformSplat}, a novel framework for rigidity-aware deformation of 3D Gaussians using only a single target image.
\item We present Gaussian-to-Pixel Matching strategy that connects the 3D Gaussian representation with the 2D target image to guide deformation.
\item We propose Rigid Part Segmentation method, which preserves the original geometry by detecting rigid regions and constraining their deformation.
\item Our method shows superior performance in single image Gaussian deformation and extends to applications such as frame interpolation and interactive manipulation.
\end{itemize}
\begin{figure*}
    \centering
    \includegraphics[width=0.8\linewidth]{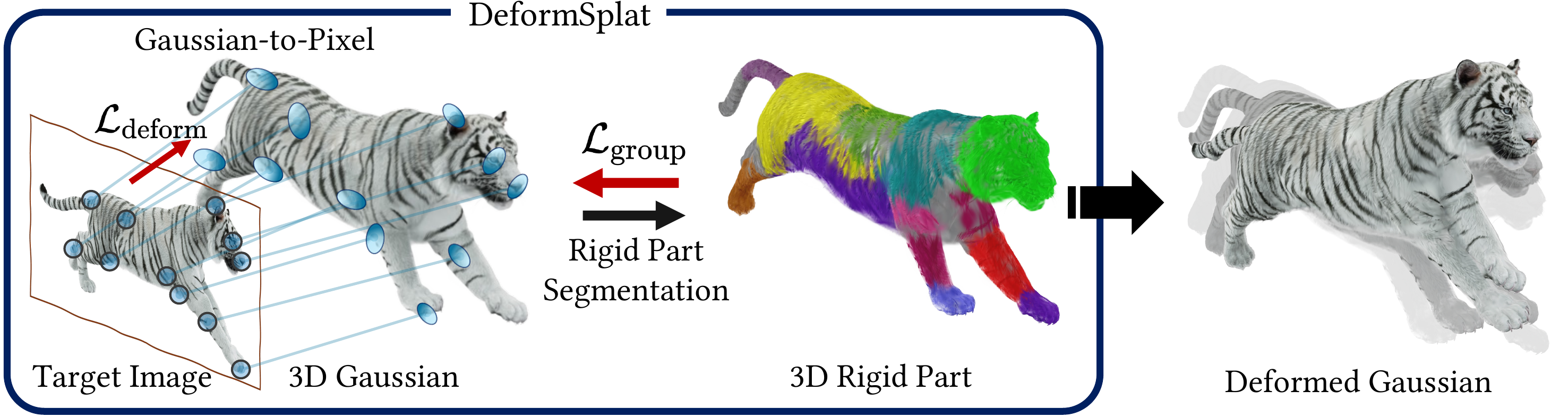}
    \caption{
    \textit{Overview of \texttt{DeformSplat}.} It first establishes correspondences between the 2D target image and 3D Gaussian. Then, Rigid Part Segmentation explicitly identifies rigid regions. By combining two methods, our approach reconstructs the deformed Gaussian, which can be rendered from novel views.
    }    
    \label{fig:full_pipe}
\end{figure*}
\section{Related Work}

\subsection{Dynamic Reconstruction}
Dynamic reconstruction is the task of recovering time-varying 3D geometry, including motion and non-rigid deformations, in real-world.
One influential line of work~\cite{pumarola2021dnerf,li2022neural,park2023temporal,park2021hypernerf,guo2023forward} is based on NeRF~\cite{mildenhall2021nerf}.
D-NeRF~\cite{pumarola2021dnerf} is a representative extension of NeRF that introduces time as an additional input, enabling dynamic reconstruction and partial modeling of non-rigid motion.
However, its MLP-based implicit representation results in slow processing and limited control over localized dynamics.

Recent works~\cite{wu20244dgs,huang2024scgs,lu2024gags,yang2024deformable,luiten2024dynamic,yang2023real} have focused on 3DGS~\cite{kerbl20233dgs}, an explicit representation using 3D Gaussians that enables fast training, real-time For instance, 4D-GS~\cite{wu20244dgs} deforms a fixed set of canonical Gaussians over time via a learned deformation field, enabling real-time rendering.
SC-GS~\cite{huang2024scgs} controls motion using a small number of sparse control points, allowing efficient and editable deformation with fewer parameters.
These methods leverage the strengths of explicit representations in terms of editability and computational speed.
However, their reliance on continuous multi-view video limits their practicality in real-world settings, in which such data is often difficult to obtain.

To mitigate the difficulty of real-world data acquisition, recent work has investigated few-view and single-view dynamic reconstruction. Few-view methods such as NPG~\cite{das2024npg} and MAGS~\cite{guo2024mags} adopt low-rank bases or optical flow to better capture motion under limited viewpoints. In the single-view case, approaches like Shape of Motion~\cite{wang2024shape} and MoSCA~\cite{lei2025mosca} leverage priors such as depth and tracking models, while CUT3R~\cite{wang2025continuous} reconstructs camera pose and dynamics in a feed-forward manner. More recent methods, including MegaSAM~\cite{li2025megasam} and ViPE~\cite{huang2025vipe}, combine video depth prediction with bundle adjustment for fast dynamic reconstruction.
However, one major drawback of these methods is that they depend on continuous video.
In particular, limited frame-to-frame continuity, such as with low or inconsistent frame rates, leads to a significant decline in reconstruction performance.
These challenges collectively point to the need for a framework that can explicitly perform 3D Gaussian deformation from a single image.

\subsection{3D Gaussian Editing}

3D editing refers to the interactive modification of 3D representations based on user input.
Recent research highlights 3DGS for enabling fast, photorealistic rendering and intuitive editing through its explicit representation.

Text-driven 3D editing has emerged as a promising direction within 3D Gaussian editing, where user-provided text prompts are used to modify 3DGS representations.
Recent works, such as GaussCtrl~\cite{wu2024gaussctrl}, GaussianEditor~\cite{chen2024gaussianeditor}, GSEdit~\cite{palandra2024gsedit}, and GSEditPro~\cite{sun2024gseditpro}, utilize 2D diffusion models conditioned on text prompts to modify the appearance of 3D Gaussians.
To enhance editing fidelity and usability, these methods incorporate additional techniques, such as depth-aware consistency~\cite{wu2024gaussctrl}, semantic region tracking~\cite{chen2024gaussianeditor}, fast object-level modification~\cite{palandra2024gsedit}, and attention-guided localization~\cite{sun2024gseditpro}.
This research direction enables intuitive and diverse editing without requiring expert skills.
However, the ambiguity of natural language can lead to varied interpretations, causing inconsistent results due to the limitations of 2D diffusion models.

Complementary to text-based methods, image-driven 3D editing enables intuitive modification of 3DGS through visual inputs, allowing users to express their intent more clearly.
Representative works, such as ReGS~\cite{mei2024regs}, ICE-G~\cite{jaganathan2024iceg}, and ZDySS~\cite{saroha2025zdyss}, enable appearance editing based on reference images.
Specifically, they address texture underfitting~\cite{mei2024regs}, enable localized appearance transfer~\cite{jaganathan2024iceg}, and support zero-shot stylization~\cite{saroha2025zdyss}. 
These works allow intuitive editing by directly linking 2D inputs to 3D output, but current methods focus only on appearance, highlighting the need for geometry-level control.
GESI~\cite{luo2024gesi} addresses this by modifying 3D Gaussians based on a reference image and camera pose, following a principle of “what you see is what you get”.
However, the lack of explicit separation between rigid and deformable regions makes it difficult to preserve structural integrity and geometric consistency during deformation.
To address previous limitations, we propose a single-image deformation framework that preserves the object's structural integrity by explicitly separating rigid and non-rigid components.
This enables stable dynamic reconstruction without relying on multi-view or video input, making it practical for real-world use.

\section{Method}
\label{sec:method}
\subsection{Task Overview}
\label{sec:task_overview}

Given a pre-reconstructed 3D Gaussian and a single target image depicting a deformed object, our goal is to deform the 3D Gaussian to accurately match the deformation observed in the target image. 
At the same time, we aim to preserve the original geometry of the initial 3D Gaussian. An overview
of our method is shown in Fig.~\ref{fig:full_pipe}.
Formally, let the 3D Gaussian $\mathcal{G} = \{\mu_i, q_i, s_i, \alpha_i, sh_i\}$ denote a set of Gaussians, each defined by its center position $\mu_i \in \mathbb{R}^3$, quaternion $q_i\in \mathbb{R}^4$, scale $s_i \in \mathbb{R}^3$, opacity $\alpha_i\in \mathbb{R}^1$, and spherical harmonic coefficients $sh_i\in \mathbb{R}^{48}$.
This Gaussian representation is initially reconstructed from multiple views captured at an earlier time step. 
The target deformation is provided as a single RGB image $\mathcal{I}_{\text{target}}$, 
without any explicit 3D information, such as depth or camera pose. To efficiently find the underlying deformation, we only optimize location $\mu_i$ and rotations $q_i$ in order to align with the deformation depicted in the target image  $\mathcal{I}_{\text{target}}$.

Our task is particularly difficult due to limited input conditions.
Conventional dynamic reconstruction approaches~\cite{wu20244dgs,lu2024gags,huang2024scgs} typically rely on abundant multi-view images or continuous temporal data to robustly model deformation.
In contrast, our method is restricted to supervision from just a single image, complicating accurate deformation guidance.
Furthermore, even though multiple camera poses are known from the initial Gaussian reconstruction, the exact viewpoint corresponding to the target image remains unknown. This ambiguity poses additional difficulties in precisely aligning the Gaussian with the observed 2D deformation.



Under these constraints, we face two significant challenges.  
First, it is difficult to determine how each Gaussian should deform from only a single image, since this requires reliable correspondences between 3D Gaussians and 2D image.
Second, without depth or multi-view constraints, deformation can easily overfit the target image and cause distortions, even in rigid regions that should remain unchanged.  
To address these challenges, we propose two key components:
(1) Gaussian-to-Pixel Matching that selects the most overlapping viewpoint and establishes 3D-to-2D correspondences for deformation guidance, and  
(2) Rigid Part Segmentation that detects and preserves rigid regions through initialization and refinement.  
Together, these components enable accurate single-image Gaussian deformation while maintaining geometric consistency.

\begin{figure}
    \centering
    \includegraphics[width=\linewidth]{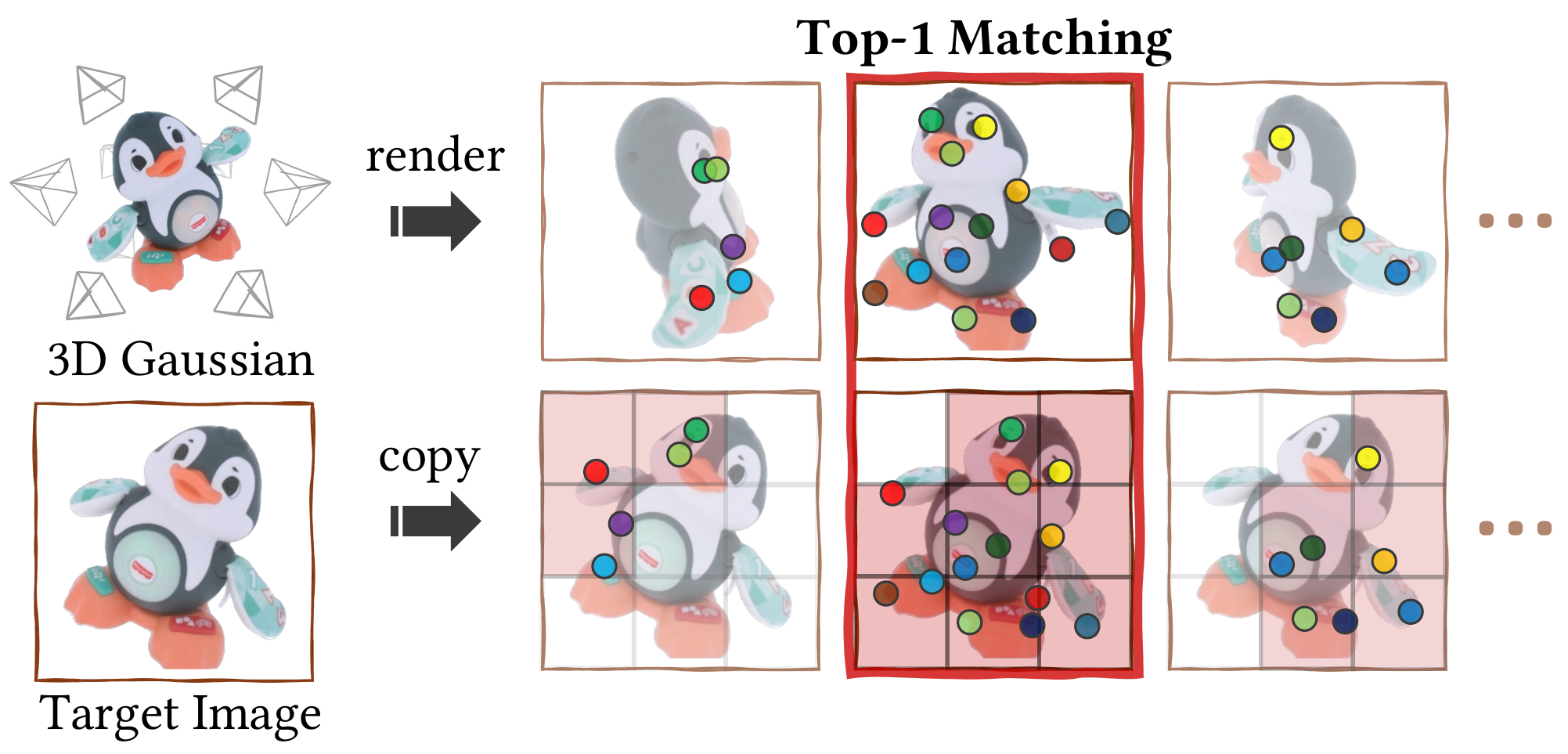}
    \caption{\textit{High-overlap image selection.} We render images from multiple cameras used initial Gaussian reconstruction. Each rendered image is matched with the target image to measure the overlap.}
    \label{fig:view_selection}
\end{figure}

\subsection{Gaussian-to-Pixel Matching}
\label{sec:gaussian_to_pixel}

Given a reconstructed 3D Gaussian and an unposed target image, a naive approach is to randomly select a camera viewpoint from those used in the initial Gaussian reconstruction. 
Then, the Gaussian can be rendered and optimized using pixel-wise loss. 
However, this strategy often fails. 
The random view may have minimal visual overlap with the target image, making pixel-wise guidance ineffective. 
Even with a large overlap, pixel-wise loss alone cannot accurately handle long-range deformation because the target image represents a deformed shape compared to the initial Gaussian.
Thus, a more precise method is required to guide deformation robustly.

To address this challenge, we propose Gaussian-to-Pixel Matching approach. 
We start by rendering multiple images from the original set of camera viewpoints used for the initial Gaussian reconstruction. 
We then apply an image matcher, RoMA~\cite{edstedt2024roma}, between each rendered image and the target image, obtaining corresponding pixel pairs $(x_p, x'_p)$, where $x_p$ represents pixels from rendered images and $x_p'$ represents corresponding pixels in the target image. 
In order to select the best viewpoint, we partition the target image into evenly spaced grids. 
For each rendered viewpoint, we count how many of these grids contain matched pixels $x'_p$.
We then select the camera viewpoint with the maximum number of matched grids, ensuring visual alignment with the target deformation. 
Fig.~\ref{fig:view_selection} illustrates this selection procedure, clearly depicting the manner in which visual overlap across multiple camera viewpoints can be quantified.

After selecting the viewpoint, we convert the pixel-to-pixel correspondences into Gaussian-to-Pixel correspondences. 
Specifically, we first evaluate the visibility of Gaussians using alpha-blended opacity $\alpha_i\prod(1 - \alpha_j)$ to exclude invisible Gaussians from the selected viewpoint. 
Among visible Gaussians, we associate each matched pixel $x_p$ with the nearest projected Gaussian center $\mu_i^{2D} = P\mu_i$, where $P$ denotes the camera projection matrix. 
Pixels sufficiently close to Gaussian projections are then replaced by the corresponding Gaussian centers $\mu_i$, establishing Gaussian-to-Pixel correspondences $(\mu_i, x_p')$.
The derived 3D-to-2D matches inherently capture the necessary directional information for guiding the Gaussians' movements. 
Leveraging this matching, we effectively guide the deformation process at a detailed level.


\begin{figure}
    \centering
    \includegraphics[width=.9\linewidth]{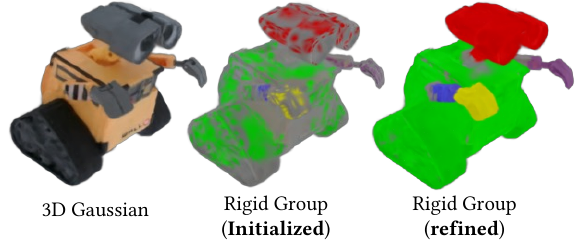}
    \caption{\textit{Example of Initialization and Refinement}. 
    Rigid groups are initialized from Gaussian-to-Pixel correspondences $(\mu_i, x'_p)$. Then, refinement step further expands these groups to cover broader regions. 
    Each color denotes an independent rigid group, while grey indicates ungrouped Gaussians.
    }
    \label{fig:rigid_seg}
\end{figure}

\subsection{Rigid Part Segmentation}
\label{sec:rigid_segmentation}

Although Gaussian-to-Pixel Matching strategy provides effective guidance for Gaussian deformation, it does not inherently guarantee preservation of the original geometry. 
%
%
To preserve the geometric of the reconstructed Gaussian during deformation, we propose two-stage rigid segmentation composed of a region-growing initialization step followed by a refinement step.

Rigid regions typically share two key properties: (1) they undergo the same rigid transformations, and (2) they exhibit strong spatial connection. 
The proposed rigid segmentation leverages these properties to robustly identify coherent rigid regions, significantly enhancing geometry preservation during deformation.
\paragraph{Rigid Part Initialization}
In the rigid initialization stage, we first utilize Perspective-n-Point (PnP) algorithm~\cite{lepetit2009ep}, which estimates a rigid transformation from 3D-to-2D correspondences. 
To identify subset Gaussians sharing similar rigid transformations, we combine the PnP algorithm with RANSAC~\cite{fischler1981random}. Specifically, PnP estimates rigid transformations from Gaussian-to-Pixel correspondences, while RANSAC robustly selects the most consistent subset of correspondences.
For simplicity, we refer to this combination as PnP-RANSAC.
Given Gaussian-to-Pixel correspondences $(\mu_i, x_p')$ obtained previously, PnP-RANSAC identifies subsets of Gaussians sharing similar rigid transformations. 
However, although sharing similar transformations is a necessary condition for rigid grouping, it alone does not guarantee spatial coherence among Gaussians. 
A meaningful rigid group should consist of Gaussians that are spatially connected.
If spatial connectivity is not enforced, distant and unrelated Gaussians may coincidentally share similar transformations. 
For instance, left and right hands might exhibit similar transformations by chance, yet they clearly should not belong to the same rigid region.

To guarantee spatial connectivity, we propose a region-growing strategy for rigid group initialization.
We begin this process by forming an initial rigid group $G$ from a randomly selected single Gaussian. 
This group is then iteratively expanded to include neighboring Gaussians using ball queries.
After each expansion, we apply PnP-RANSAC to identify inliers $G_{\text{inlier}}$ sharing a consistent rigid transformation. 
This iterative expansion and filtering continue until convergence, ensuring the rigid groups are spatially coherent and transformation-consistent. 
For detailed rigid initialization, please refer to the supplementary material.

While the rigid initialization produces spatially coherent groups, its scope is restricted to Gaussians derived from Gaussian-to-Pixel correspondences. This means that only small subsets of the rigid parts are actually initialized, as shown in Fig.~\ref{fig:rigid_seg}. To complete the rigid segmentation, we introduce a refinement stage. This process extends the segmentation to most of the Gaussians.

\paragraph{Rigid Part Refinement} The refinement stage leverages the same characteristics of rigid regions: consistent rigid transformations and spatial connectivity.
However, the key difference from the initialization is that refinement is independent of Gaussian-to-Pixel Matching and utilizes continuously updated Gaussian parameters (positions $\mu_i'$ and rotations $q_i'$) obtained during optimization. \modified{Note that the refinement is an iterative process performed jointly with rigid-aware optimization (cf. Sec. 3.4).}

In particular, we iteratively refine initially identified rigid group by enlarging it with neighboring Gaussians found via local ball queries. To evaluate whether newly added candidate Gaussians adhere consistently to the group transformation, we propose a rigidity score inspired by the As-Rigid-As-Possible (ARAP) principle~\cite{sorkine2007rigid}. Given a candidate Gaussian $\mu_i$, the rigidity score relative to an existing rigid group $G$ is computed as:
\begin{equation}
S_{\text{rigid}}(\mu_i, G) = \frac{1}{|G|}\sum_{\mu_j \in G} || {R_i^{-1}}(\mu_i - \mu_j) - {R_i'^{-1}}(\mu_i' - \mu_j')||^2,
\end{equation}
where $(\mu_i, q_i)$ and $(\mu_j', q_j')$ denote the initial and optimized Gaussian positions and rotations, respectively. $R_i$ and $R'_i$ are the rotation matrices derived from quaternion $q_i$ and $q'_i$, respectively. $|G|$ is the number of Gaussians in the group.
A small rigidity score indicates strong consistency, thus justifying the inclusion of the candidate Gaussian into the rigid group.

During each iteration, we first expand the current rigid group $G$ by identifying candidate Gaussians $\mu_i \in G_{\text{expand}}$ within a local ball-query radius. 
Each candidate Gaussian is evaluated based on its rigidity score $S_{\text{rigid}}(\mu_i, G)$.
Candidates with rigidity scores below a lower threshold $\tau_{\text{low}}$ are included in the rigid group $G$, whereas those exceeding an upper threshold $\tau_{\text{high}}$ are excluded if previously part of the group. 
Through iterative inclusion and exclusion, this refinement procedure progressively corrects and expands rigid groups, ensuring robust geometry preservation throughout deformation optimization. 
For the detailed procedure, please refer to the supplementary material.

By the combined rigid initialization and refinement steps, our Rigid Part Segmentation robustly identifies spatially coherent rigid regions, ensuring strong geometric consistency throughout the deformation process.

\subsection{Rigid-Aware Optimization}
\label{sec:rigidity_optimization}

Directly optimizing Gaussian parameters independently often leads to excessive flexibility, potentially disrupting the deformation quality. 
To effectively mitigate this issue, we adopt an anchor-based deformation representation following previous works~\cite{sumner2007embedded,huang2024scgs}.
Specifically, the deformation is represented using a sparse set of anchors, each parameterized by its position $a_k \in \mathbb{R}^3$, rotation $R_k^a \in SO(3)$ (equivalently as quaternion $q_k^a$), and translation $T_k \in \mathbb{R}^3$. 
Anchor positions are initialized by voxelizing the space and computing the average Gaussian positions within each voxel. Using these anchors, updated Gaussian positions $\mu_i'$ and rotations $q_i'$ are computed by interpolating transformations of neighboring anchors $\mathcal{N}$ as follows:
\begin{equation}
\mu_i' = \sum_{k \in \mathcal{N}} w_{ik}\left(R_k^a(\mu_i - a_k) + a_k + T_k\right), \ \   q_i' = q_i \otimes \sum_{k \in \mathcal{N}} w_{ik} q_k^a,
\end{equation}
where $\otimes$ is the production operation of quaternions and $w_{ik}$ is interpolation weight inversely proportional to distances between Gaussian $\mu_i$ and anchors $a_k$.
This sparse anchor representation significantly reduces deformation complexity, promoting smoothness and geometric coherence through localized transformations.


\paragraph{Deformation Loss.}
Using the Gaussian-to-Pixel correspondences, we define a deformation loss as follows:
\begin{equation}
\mathcal{L}_{\text{deform}} = \sum_i ||\mu_i^{2D} - x_p'||^2,
\end{equation}
which encourages Gaussian centers to move toward matched pixel locations. This approach effectively guides the deformation based on structurally meaningful matches. 

\paragraph{Rigid Group Regularization.}
Using the rigid groups from Sec.~\ref{sec:rigid_segmentation}, we introduce a group-based rigidity loss $\mathcal{L}_{\text{group}}$ to explicitly preserve geometric consistency within rigid regions. Specifically, within each rigid group $G$, we enforce consistency between the original and transformed Gaussian structures as:
\begin{equation}
\mathcal{L}_{\text{group}} = \sum_{G_k} \sum_{\mu_i,\mu_j \in G_k}|| {R_i^{-1}}(\mu_i - \mu_j) - {R_i'^{-1}}(\mu_i' - \mu_j')||^2,
\end{equation}
where $R_i$ and $R_i'$ denote rotation matrices derived from original and updated Gaussian rotations $q_i$ and $q_i'$, respectively. 

\paragraph{ARAP Regularization.}
While $\mathcal{L}_{\text{group}}$ explicitly preserves geometry within rigid regions, non-rigid regions also require regularization to ensure smooth and natural deformation. To achieve this, we apply ARAP regularization between neighboring anchors as follows:
\begin{equation}
\mathcal{L}_{\text{arap}} = \sum_{a_i}\sum_{k \in \mathcal{N}}|| R_i^a(a_i - a_k) - (a_i' - a_k')||^2,
\end{equation}
where $a_i' = a_i + T_i$ is the updated anchor position, and $R_i^a$ is the rotation at anchor $a_i$. This ARAP loss promotes local rigidity among anchors, resulting in coherent deformation transitions.

\paragraph{RGB Loss.}
To ensure visual alignment with the target deformation, we employ a photometric RGB loss aligning the rendered image $\mathcal{I}_{\text{render}}$ with the target image $\mathcal{I}_{\text{target}}$:
\begin{equation}
\mathcal{L}_{\text{rgb}} = ||\mathcal{I}_{\text{render}} - \mathcal{I}_{\text{target}}||^2.
\end{equation}


\begin{table*}[t]
\caption{\textit{Quantitative result on the Diva360 and DFA datasets.} Best results are indicated in \textbf{bold} and second-best results are \underline{underlined}.}
\centering
\resizebox{1.35\columnwidth}{!}{
\begin{tabular}{lcccccc}
\toprule
\multirow{2}{*}{Method} & \multicolumn{3}{c}{Diva360} & \multicolumn{3}{c}{DFA} \\ \cmidrule(lr){2-4} \cmidrule(lr){5-7}
 & PSNR$\uparrow$ & SSIM$\uparrow$ & LPIPS$\downarrow$ & PSNR$\uparrow$ & SSIM$\uparrow$ & LPIPS$\downarrow$ \\ 
\midrule
3DGS~\cite{kerbl20233dgs} & 21.28 & 0.900 & 0.098 & 19.48 & 0.866 & 0.118 \\
\modified{DROT~\cite{xing2022differentiable}}
& \modified{21.08} & \modified{0.914} & \modified{0.086} 
& \modified{17.64} & \modified{0.872} & \modified{0.119}  \\
SC-GS~\cite{huang2024scgs} & 22.20 & 0.910 & 0.097 & 19.49 & 0.867 & 0.116 \\
4DGS~\cite{wu20244dgs} & 19.93 & 0.913 & 0.100 & 14.56 & 0.856 & 0.204 \\
\modified{3DGStream~\cite{wu20244dgs}}
& \modified{22.57} & \modified{{0.928}} & \modified{0.088}
& \modified{\underline{20.16}} & \modified{{0.886}} & \modified{\underline{0.100}} \\
GESI~\cite{luo2024gesi} & \underline{22.71} & 0.897 & 0.086 & 18.54 & 0.876 & 0.127 \\
GESI ($\mu, q$)~\cite{luo2024gesi} & 22.53 & \underline{0.924} & \underline{0.078} & {20.05} &  \underline{0.888} &  \underline{0.100} \\ 
\midrule
Ours & \textbf{26.84} & \textbf{0.955} & \textbf{0.050} & \textbf{21.81} & \textbf{0.897} & \textbf{0.091} \\
\bottomrule
\end{tabular}
}
\label{tab:combined_comparison}
\end{table*}

\begin{figure*}
    \centering
    \includegraphics[width=.86\linewidth]{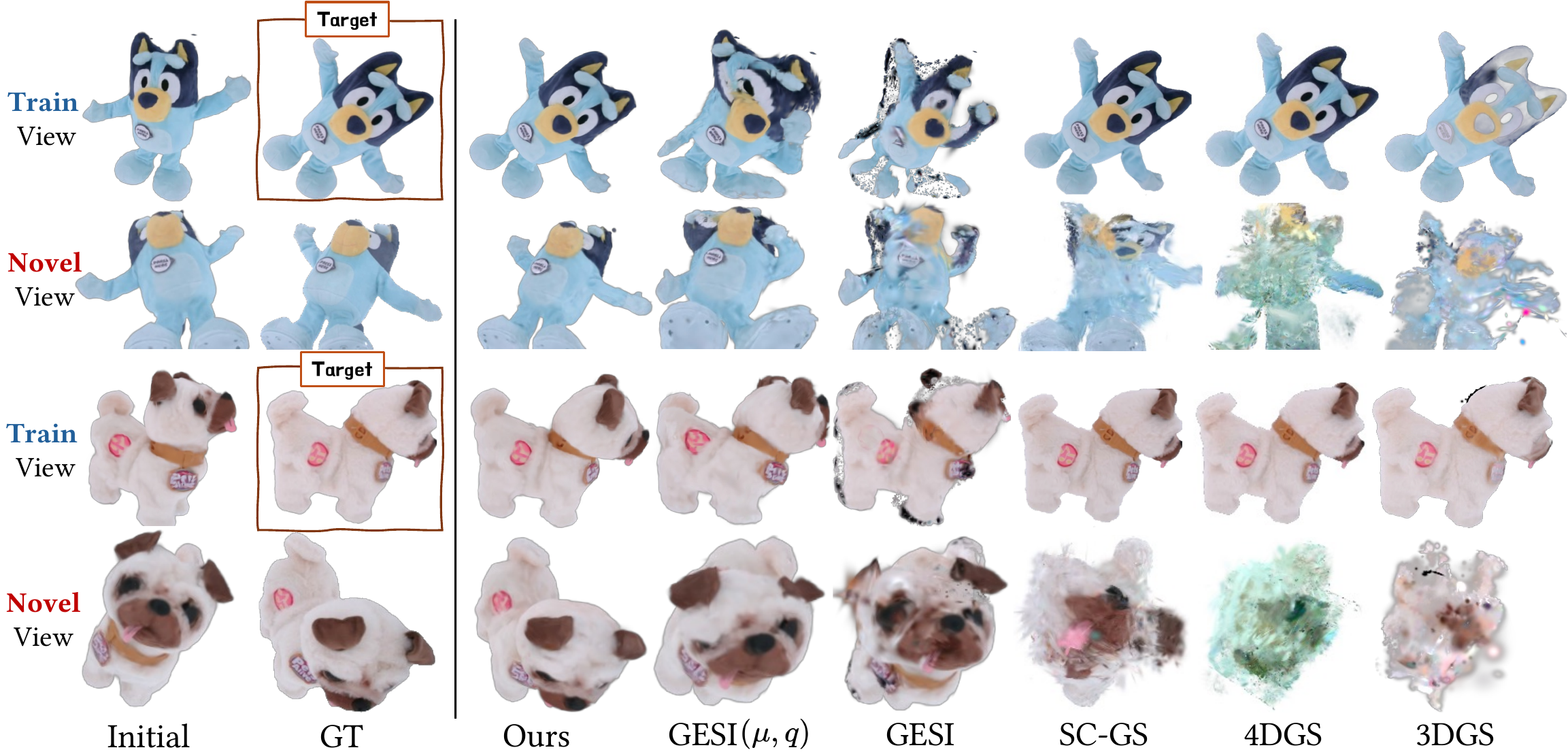}
    \caption{\textit{Qualitative comparison on the Diva360 dataset.}  The target images are highlighted with brown boxes in the second column.}
    \label{fig:diva_result}
\end{figure*}

\paragraph{Total Optimization Loss.}
Combining these terms, we obtain our total optimization objective:
\begin{equation}
\mathcal{L}_{\text{total}} 
= \lambda_{\text{deform}} \mathcal{L}_{\text{deform}} 
+ \lambda_{\text{group}} \, \mathcal{L}_{\text{group}} 
+ \lambda_{\text{arap}} \, \mathcal{L}_{\text{arap}} 
+ \lambda_{\text{rgb}} \, \mathcal{L}_{\text{rgb}}.
\end{equation}
Here, each $\lambda$ denotes a hyperparameter that balances the corresponding loss term.
The unified optimization scheme simultaneously guides accurate deformation, preserves rigid region geometry, and ensures visual consistency.

\paragraph{Smooth Motion Interpolation.} After optimization, we further introduce post-processing for smooth interpolation. Specifically, we define an interpolation loss as follows:
\begin{equation}
\mathcal{L}_{\text{inter}} = \sum_i\left(||R_i^a - \hat{R}_i^a|| + ||T_i^a - \hat{T}_i^a||\right) + \lambda_{\text{inter}} (\mathcal{L}_{\text{group}} + \mathcal{L}_{\text{arap}}),
\end{equation}
where $(R_i^a, T_i^a)$ and  $(\hat{R}_i^a, \hat{T}_i^a)$ denote initial and optimized anchor transformations. $\lambda_{\text{inter}}$ is decaying hyperparameter to ensure convergence.
As $(R_i^a, T_i^a)$ gradually approaches $(\hat{R}_i^a, \hat{T}_i^a)$, we achieve smooth motion transitions that preserve geometric consistency, resulting in visually pleasing deformation outcomes.

\begin{figure*}
    \centering
    \includegraphics[width=1\linewidth]{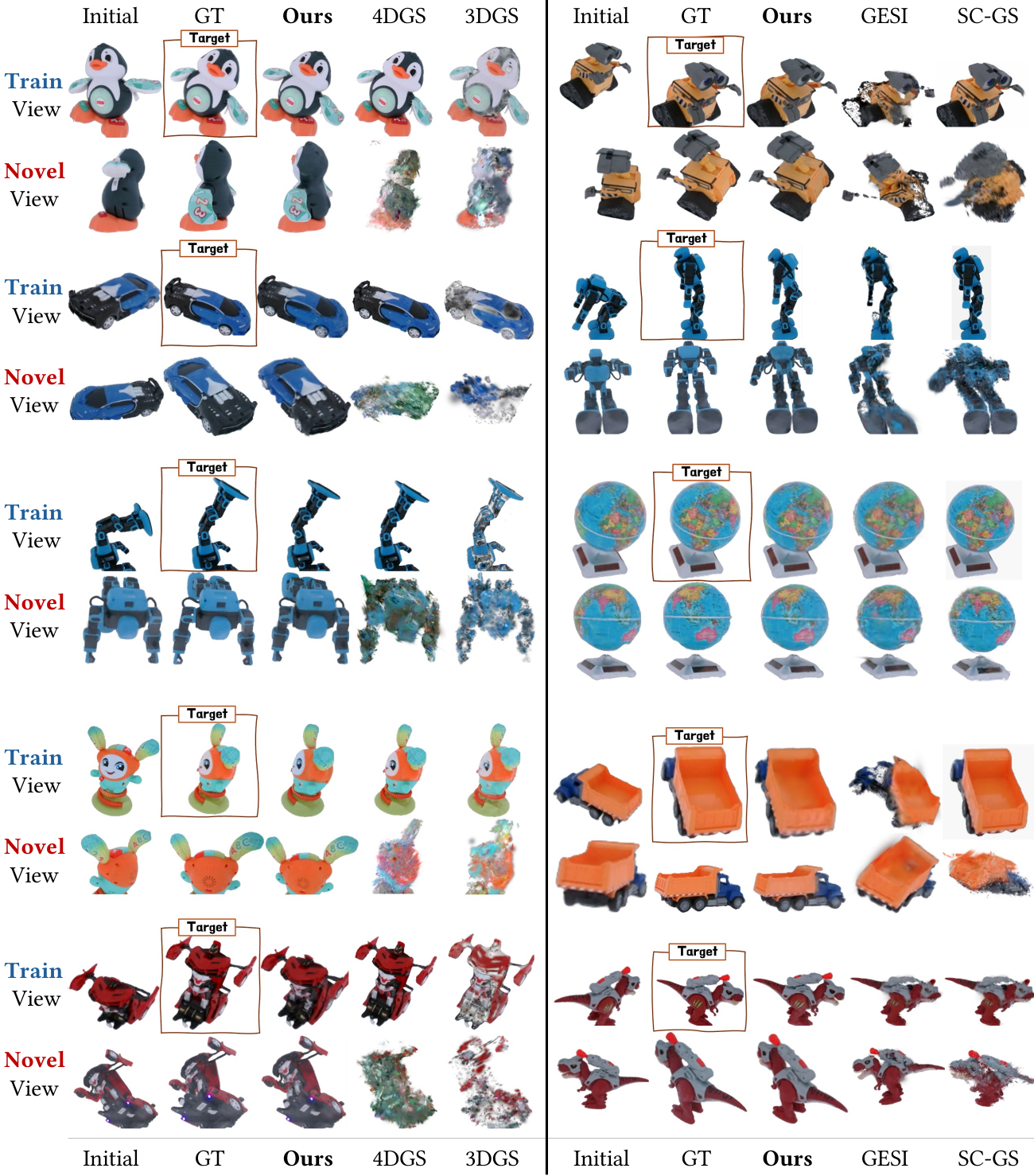}
    \caption{\textit{Diverse result on the Diva360 dataset.} The target images are highlighted with brown boxes in the second column.}
    \label{fig:diva_comparison}
\end{figure*}

\begin{figure*}
    \centering
    \includegraphics[width=0.9\linewidth]{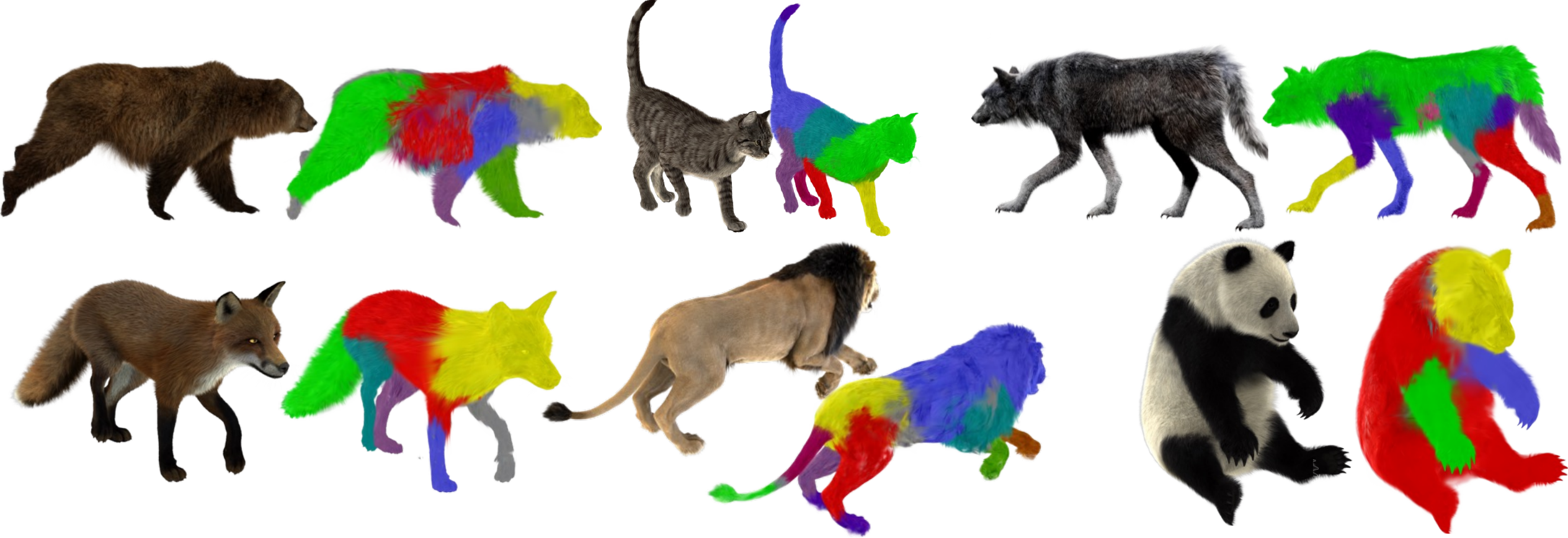}
    \caption{\textit{Rigid group visualization on the DFA dataset.} Grey region refers ungrouped Gaussian.}
    \label{fig:dfa_rigid}
\end{figure*}

\begin{figure*}
    \centering
    \includegraphics[width=1\linewidth]{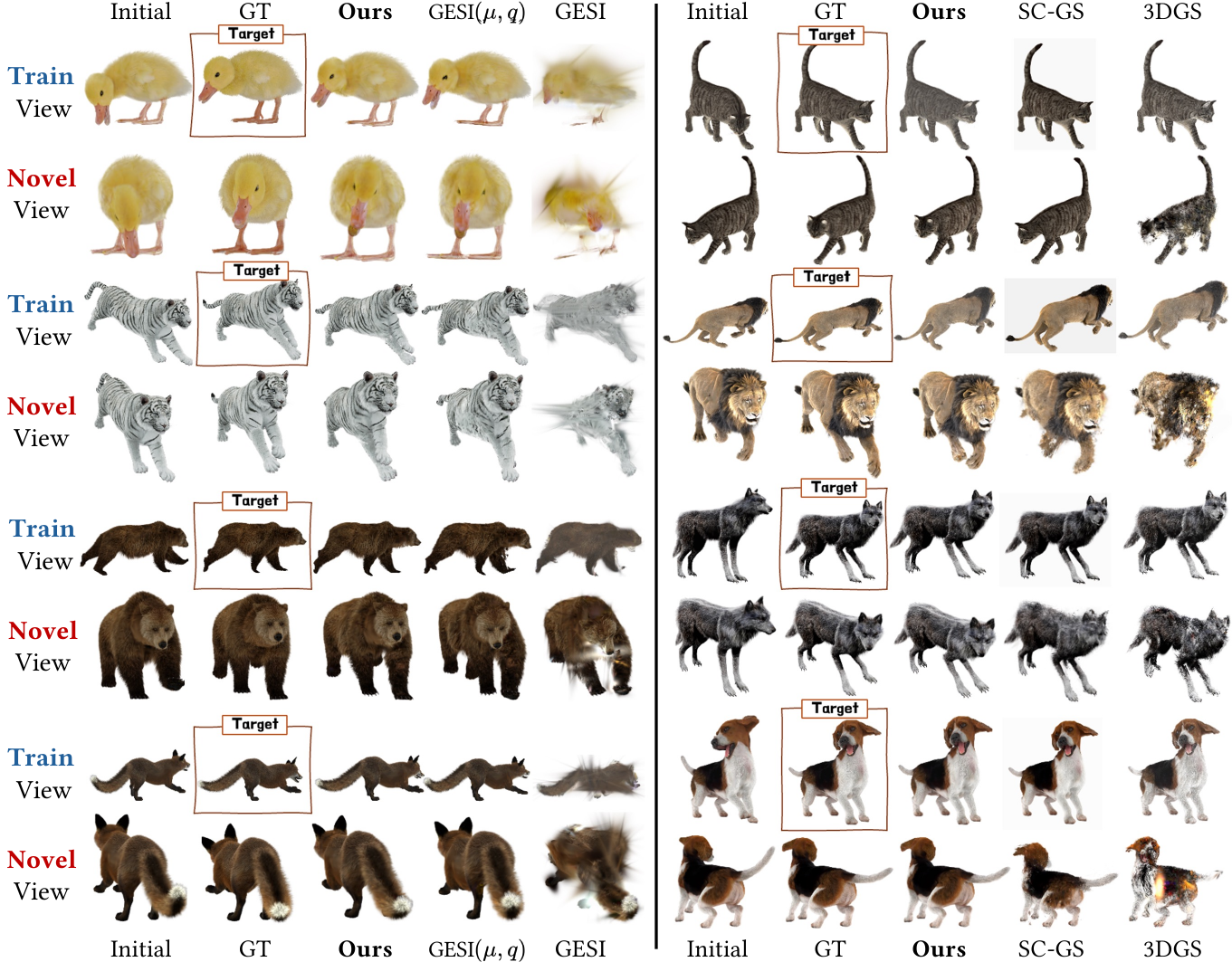}
    \caption{\textit{Diverse result on the DFA dataset.} The target images are highlighted with brown boxes in the second column.}
    \label{fig:dfa_comparison}
\end{figure*}

\section{Experiment}
\label{sec:experiment}

\subsection{Experiment Setting}

\paragraph{Datasets.}
We evaluate ours on two multi-view video datasets: diverse moving object sequences in the Diva360 dataset~\cite{lu2024diva} and the synthetic Dynamic Furry Animal (DFA) dataset~\cite{luo2022artemis}. The Diva360 dataset captures various dynamic objects from multiple views in a $360^\circ$ configuration and comprises 21 sequences. Among them, we exclude two ``Plasma Ball'' sequences since they show only light changes and do not exhibit any deformation. The synthetic DFA dataset, generated from motion capture data, includes 25 sequences depicting animated animal movements.

\modified{
For each of the $N$ video sequences in each dataset, we select two distinct timesteps. 
For the first timestep, we select a moment where the object is fully visible without occlusions. 
This enables accurate initial reconstruction of the Gaussian model using images from multiple views. 
The second timestep, chosen to represent the target deformation state, contains noticeable deformation. 
The target deformation is supervised using only a single viewpoint image from the second timestep. 
The remaining viewpoint images from the second timestep are used as ground-truth images for evaluation. 
These two timesteps are manually selected to ensure noticeable deformation, minimal occlusion, and sufficient visual consistency.
The selected data samples can be founded through our released code.
}

\paragraph{Baselines.}
To validate our approach, we compare \texttt{DeformSplat} with established baseline methods. 
Specifically, we compare ours with 3DGS~\cite{kerbl20233dgs}, which directly optimizes reconstructed Gaussians using pixel-wise RGB losses (L1 and SSIM) from a single target image. 
\modified{
We also include a comparison with 3DGS that is optimized using the optical transport RGB loss, called DROT~\cite{xing2022differentiable}, which enables more robust, long-range comparisons rather than simple pixel-wise differences. 
Additionally, we compare against two dynamic Gaussian reconstruction methods (4DGS~\cite{wu20244dgs}, SC-GS~\cite{huang2024scgs}) and a streamable Gaussian method (3DGStream~\cite{sun20243dgstream}).
}
Lastly, we evaluate GESI~\cite{luo2024gesi}, a Gaussian editing method designed for single-image input, in two variants: one optimizing all Gaussian parameters, and another selectively optimizing only Gaussian positions $\mu$ and rotations $q$. 
The selective tuning of ($\mu, q$) parameters aims to represent deformation more explicitly, ensuring a fairer comparison with our method.
Since there is no publicly available implementation for GESI, we implement it following the details provided in the original paper.

\paragraph{Evaluation Metrics and Camera Alignment.}
We quantitatively evaluate deformation accuracy using three metrics: Peak Signal-to-Noise Ratio (PSNR), Structural Similarity Index Measure (SSIM), and Learned Perceptual Image Patch Similarity (LPIPS)~\cite{zhang2018unreasonable}. 
Since our approach selects one camera viewpoint based on visual overlap, the camera pose might not be aligned with the global coordinate. Therefore, after optimization, the camera pose and final Gaussian are rotated and translated to align with the ground-truth camera poses. 
This enables fair quantitative and qualitative evaluation. All baseline methods directly use these ground-truth camera poses, ensuring consistency across comparisons.


\subsection{Comparison with Baselines}

\paragraph{Qualitative Comparison.}
Fig.~\ref{fig:diva_result}, Fig.~\ref{fig:diva_comparison} and Fig.~\ref{fig:dfa_comparison} present qualitative results comparing our method with the baseline approaches on the Diva360 and DFA datasets, respectively. \texttt{DeformSplat} successfully reconstructs accurate and visually consistent deformation from the single-target-image input, achieving high visual similarity to the ground-truth reference images. In contrast, baseline methods demonstrate notable visual artifacts and geometric distortions when rendered from viewpoints not observed during optimization. Specifically, 3DGS, 4DGS, and SC-GS overly rely on pixel-level color losses, leading to insufficiently accurate deformations. GESI, on the other hand, fails to produce accurate deformations due to occasional inaccuracies in its long-range matching via DROT.

\paragraph{Quantitative Comparison.}
Table~\ref{tab:combined_comparison} summarizes quantitative evaluations. \texttt{DeformSplat} significantly outperforms baseline methods, setting a new SOTA performance standard. On Diva360, \texttt{DeformSplat} achieves an average PSNR increase of 4.1 compared to the next best-performing baseline. On the DFA dataset, we similarly observe a PSNR improvement of 1.8. 
These results confirm the robust capability of \texttt{DeformSplat} to reconstruct detailed object deformations from minimal supervision accurately.

Regarding the performance of other baselines, GESI demonstrates the second-best performance on the Diva360 dataset, following our method, largely due to its ability to provide long-range guidance. On the DFA dataset, 3DGStream ranks just below our method, benefiting from its efficient optimization of Gaussians with fewer iterations compared to other baselines. In contrast, 4DGS shows the lowest performance in both dataset. This is because 4DGS encodes both geometry and color into the same implicit function, causing the color to change significantly when the geometry is updated during the optimization. 



\begin{table}[t]
\caption{\textit{Ablation study of each component on  Diva360 dataset.}}
\centering
\resizebox{.95\linewidth}{!}{
\begin{tabular}{lccc}
\toprule
Method Variant & PSNR$\uparrow$ & SSIM$\uparrow$ & LPIPS$\downarrow$ \\
\midrule
w/o $L_{\text{deform}}$, w/o $L_{\text{group}}$ & 21.73 & 0.917 & 0.082 \\
w/o $L_{\text{group}}$ & 24.36 & 0.942 & 0.071 \\
w/o region-growing initialize & 25.25 & 0.946 & 0.067 \\
w/o rigid refinement & 25.79 & 0.949 & 0.061 \\
full pipeline (ours) & \textbf{26.84} & \textbf{0.955} & \textbf{0.050} \\
\bottomrule
\end{tabular}
}
\label{tab:ablation}
\end{table}

\begin{figure}
    \centering
    \includegraphics[width=0.8\linewidth]{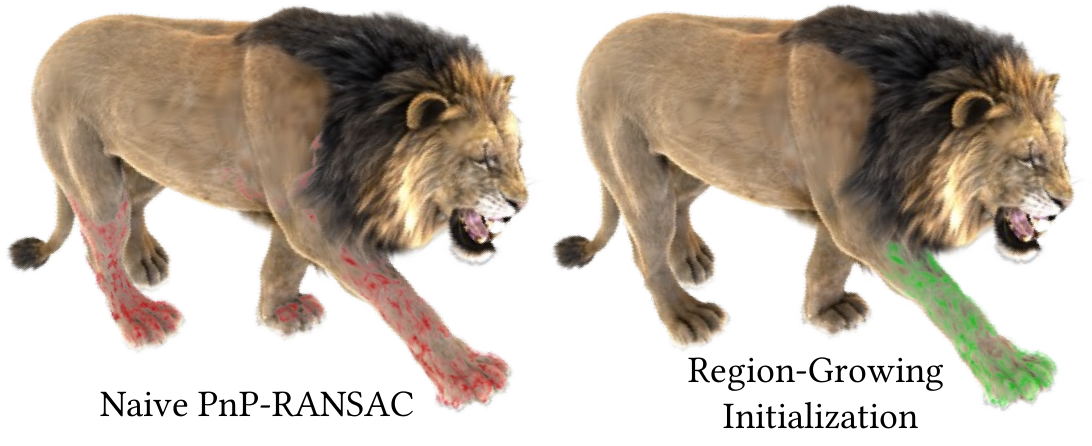}
    \caption{\textit{Ablation on Region-Growing Rigid Clustering.} Naive PnP-RANSAC yields a disconnected rigid group, while our region-growing method produces a spatially coherent group.}
    \label{fig:ablation}
\end{figure}

\subsection{Ablation Study}
We conduct an ablation study to evaluate the contribution of each component in our framework. 
As summarized in Table~\ref{tab:ablation}, removing or replacing individual modules leads to noticeable performance drops, confirming the importance of each design choice.

\paragraph{Deformation Loss.}
The absence of our deformation loss significantly degrades deformation quality. Without this structural guidance, deformation relies solely on pixel-wise color information, resulting in substantial performance degradation.

\paragraph{Rigid Group Loss.}
Removing the rigid group loss notably reduces deformation quality. This loss explicitly preserves geometry within rigid regions, highlighting its crucial role in maintaining geometry during deformation.

\paragraph{Region-Growing Initialization.}
Substituting our region-growing initialization with naive PnP-RANSAC initialization decreases deformation quality. Qualitatively, as shown in Fig.~\ref{fig:ablation}, naive PnP-RANSAC produces disconnected rigid groups, while our region-growing method effectively enforces spatial coherence.

\paragraph{Rigid Refinement.}
Removing the rigid refinement reduces deformation accuracy. This refinement iteratively updates rigid segmentation using optimized Gaussian parameters, correcting initial segmentation errors. Without it, the initial segmentation can be biased or have some errors.





\section{Application}
\label{sec:application}

\subsection{Multi-Frame Interpolation}

Given the capability of \texttt{DeformSplat} to reconstruct deformation from a single image, our framework naturally generalizes to frame interpolation for generating smooth video sequences. Specifically, when provided with an initial Gaussian reconstruction and multiple target frames, \texttt{DeformSplat} sequentially applies deformation in an autoregressive manner, using the deformed Gaussian from the previous frame as the input for the next.

\begin{figure}
    \centering
    \includegraphics[width=0.9\linewidth]{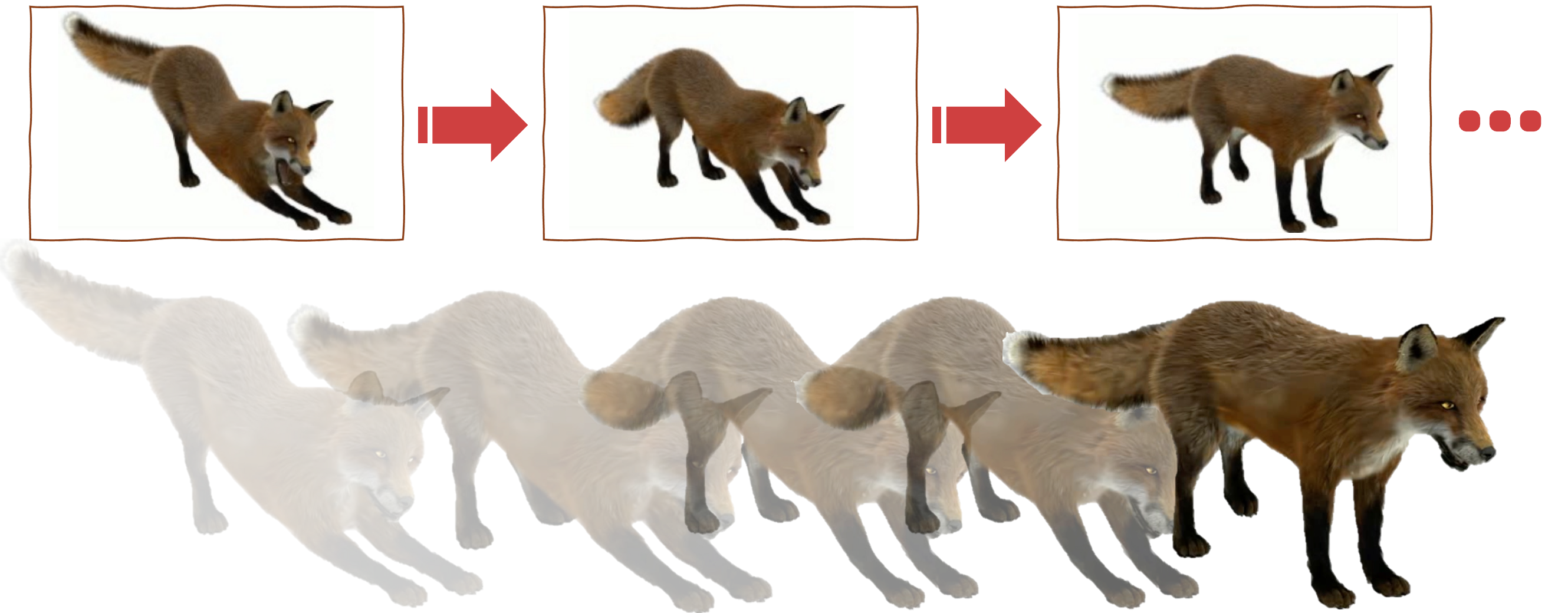}
    \caption{\textit{Multi-Frame Interpolation.} \texttt{DeformSplat} naturally extends to sequential tasks, enabling smooth frame interpolation. Please refer supplementary video for detailed examples.}
    \label{fig:frame_interpolation}
\end{figure}

As illustrated in Fig.~\ref{fig:frame_interpolation}, the resulting interpolated frames exhibit smooth transitions, accurately preserving temporal coherence and dynamic realism for various objects.
These interpolation results highlight \texttt{DeformSplat}’s capability beyond single-frame deformation, suggesting promising extensions into practical applications such as video content creation and dynamic scene generation.

\subsection{Rigid-Aware Manipulation}

Leveraging the explicit rigid segmentation and Gaussian-to-Pixel correspondences established by our framework, \texttt{DeformSplat} enables intuitive and precise interactive manipulation of 3D Gaussian objects. Specifically, users can perform direct manipulation by dragging pixels on a target image, guiding corresponding Gaussians effectively through our deformation loss. Simultaneously, our rigid group loss and ARAP regularization preserve structural integrity, ensuring natural and physically plausible transformations.

\begin{figure}
    \centering
    \includegraphics[width=\linewidth]{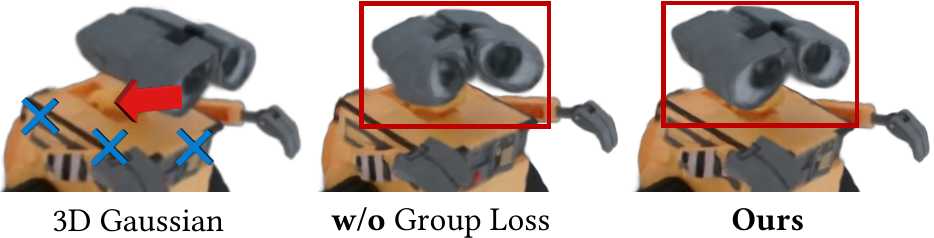}
    \caption{\textit{Interactive Manipulation.} Our rigidity-aware group loss effectively preserves underlying geometry, facilitating more natural and intuitive interactive manipulation of 3D objects.}
    \label{fig:rigid_manipulation}
\end{figure}

As shown in Fig.~\ref{fig:rigid_manipulation}, user-defined manipulations produce smooth, coherent deformations. Notably, rigid regions maintain their structural integrity with minimal distortion, while adjacent non-rigid regions deform flexibly and naturally. This rigid-aware characteristic makes \texttt{DeformSplat} highly suitable for various interactive applications, including video content editing and object manipulation in virtual reality environments.

\section{Limitations and Future Work}
While \texttt{DeformSplat} achieves SOTA results for single image-guided 3D Gaussian deformation, it still has three key limitations that we aim to address in future work.

\paragraph{Robust Matching for Dynamic Object.} The performance of our method is highly dependent on the quality of image matching. The image matcher we use, RoMA, is trained on static datasets, which sometimes leads to inaccurate correspondences when applied to dynamic objects. This can result in incorrect deformations. Improving the robustness of image matching for dynamic content is an important direction for future research.
\paragraph{Handling Fully Flexible Object.} While our approach performs well on semi-rigid deformations, it struggles with highly non-rigid objects such as clothing or fluids. This limitation arises from our assumption that the target object contains rigid components. Developing alternative strategies tailored to fully flexible objects will be essential for broadening the applicability of our method.
\paragraph{Handling Color Change.} Since \texttt{DeformSplat} only optimizes $\mu, q$ parameters, it does not handle color changes during deformation. When we attempted to optimize color jointly with geometry, color of 3D Gaussians tended to overfit, resulting in unrealistic appearances. Future work will focus on incorporating regularization strategies that enable consistent and natural color adaptation throughout deformation, without overfitting.

\section{Conclusion}

In this work, we introduced \texttt{DeformSplat}, a novel framework that reconstructs deformations of 3D Gaussians from a single image. Specifically, we proposed Gaussian-to-Pixel Matching that bridges two distinct data representations to accurately guide deformation. Additionally, we introduced Rigid Part Segmentation, a two-stage method consisting of initialization and refinement, designed to preserve original geometric structures. These techniques effectively handle long-range deformation and geometric preserving inherent in single image deformation. The experiment shows that our method significantly outperforms existing approaches and extends to diverse applications.

\begin{acks}
This work was supported by Institute of Information \& communications Technology Planning \& Evaluation~(IITP) grant funded by the Korea government~(MSIT) (No.RS-2025-25442149, LG AI STAR Talent Development Program for Leading Large-Scale Generative AI Models in the Physical AI Domain, No.RS-2025-25442824, AI Star Fellowship Program (UNIST) and No.RS-2020-II201336, Artificial Intelligence Graduate School Program (UNIST)).
\end{acks}


\bibliographystyle{ACM-Reference-Format}
\bibliography{ref}

\clearpage
\appendix









\twocolumn[{%
  \begin{center}
    \Huge \text{Supplementary Materials for}\\[8pt]
    \Huge \textit{``Rigidity-Aware 3D Gaussian Deformation from a Single Image''}
    \vspace{18pt}
  \end{center}
}]

\begin{algorithm}
    \caption{Region-Growing Rigid Initialization}
    \label{alg:region_growing}
    \begin{algorithmic}[1]
    \State $\mathcal{U} \gets {\mu_i}$ (unlabeled Gaussians)
    \While{$\mathcal{U}$ is not empty}
    \State $\mu_{\text{seed}} \gets$ random selection from $\mathcal{U}$
    \State Initialize group: $G \gets {\mu_{\text{seed}}}$
    \Repeat
    \State $G_{\text{expand}} \gets \text{BallQuery}(G)$
    \State $G_{\text{inlier}}, G_{\text{outlier}} \gets$ PnP-RANSAC($G_{\text{expand}}$)
    \If{$|G_{\text{inlier}}| \leq |G|$}
    \State Break
    \Else
    \State $G \gets G_{\text{inlier}}$
    \EndIf
    \Until{Group size converges}
    \State $\mathcal{U} \gets \mathcal{U} - G_{\text{expand}}$
    \If{$|G| \geq |G|_{\text{min}}$}
        \State $\mathcal{G}_{\text{rigid}} \gets \mathcal{G}_{\text{rigid}} \cup \{G\}$
    \EndIf
    \EndWhile
\end{algorithmic}
\end{algorithm}

\begin{algorithm}
    \caption{Rigid Group Refinement}
    \label{alg:rigid_refinement}
    \begin{algorithmic}[1]
    \For{$G$ in $\mathcal{G}_{\text{rigid}}$}
    \State $G_{\text{expand}} \gets \text{BallQuery}(G)$
    \For{$\mu_i$ in $G_{\text{expand}}$}
    \State Compute $S_{\text{rigid}}(\mu_i, G)$
    \If{$S_{\text{rigid}}(\mu_i, G) < \tau_{\text{low}}$}
    \State Add $\mu_i$ to $G$
    \ElsIf{$S_{\text{rigid}}(\mu_i, G) > \tau_{\text{high}}$ and $\mu_i \in G$}
    \State Remove $\mu_i$ from $G$
    \EndIf
    \EndFor
    \EndFor
    \end{algorithmic}
\end{algorithm}

\section{Detail of Rigid Part Segmentation}
Rigid Part Segmentation is a procedure that preserves geometric consistency during deformation by identifying rigid groups of Gaussians sharing similar rigid transformations and spatial connectivity. It consists of two stages: an initialization stage, which identifies initial rigid groups based on Gaussian-to-Pixel correspondences, and a refinement stage, which expands and refines these groups using continuously updated Gaussian positions and rotations obtained during deformation optimization. For more details, refer to Algorithm~\ref{alg:region_growing} and Algorithm~\ref{alg:rigid_refinement}.

\section{Additional Visualization}\label{sec:supp_exp_set}
We provide two additional visualizations for experiments. Firstly, we present the rigid group result after optimization. As illustrated in Fig. 1, each rigid part is segmented according to the object's joints, demonstrating the effectiveness of our Rigid Part Segmentation. Notably, the horse toy shown in the last example is an entirely rigid object, and our method accurately identifies all areas as rigid. Secondly, we present additional results regarding interacive manipulation. As depicted in Fig. 2, the rigid group segmentation clearly maintains the geometry during manipulation, ensuring geometric consistency.

\begin{figure}
    \centering
    \includegraphics[width=\linewidth]{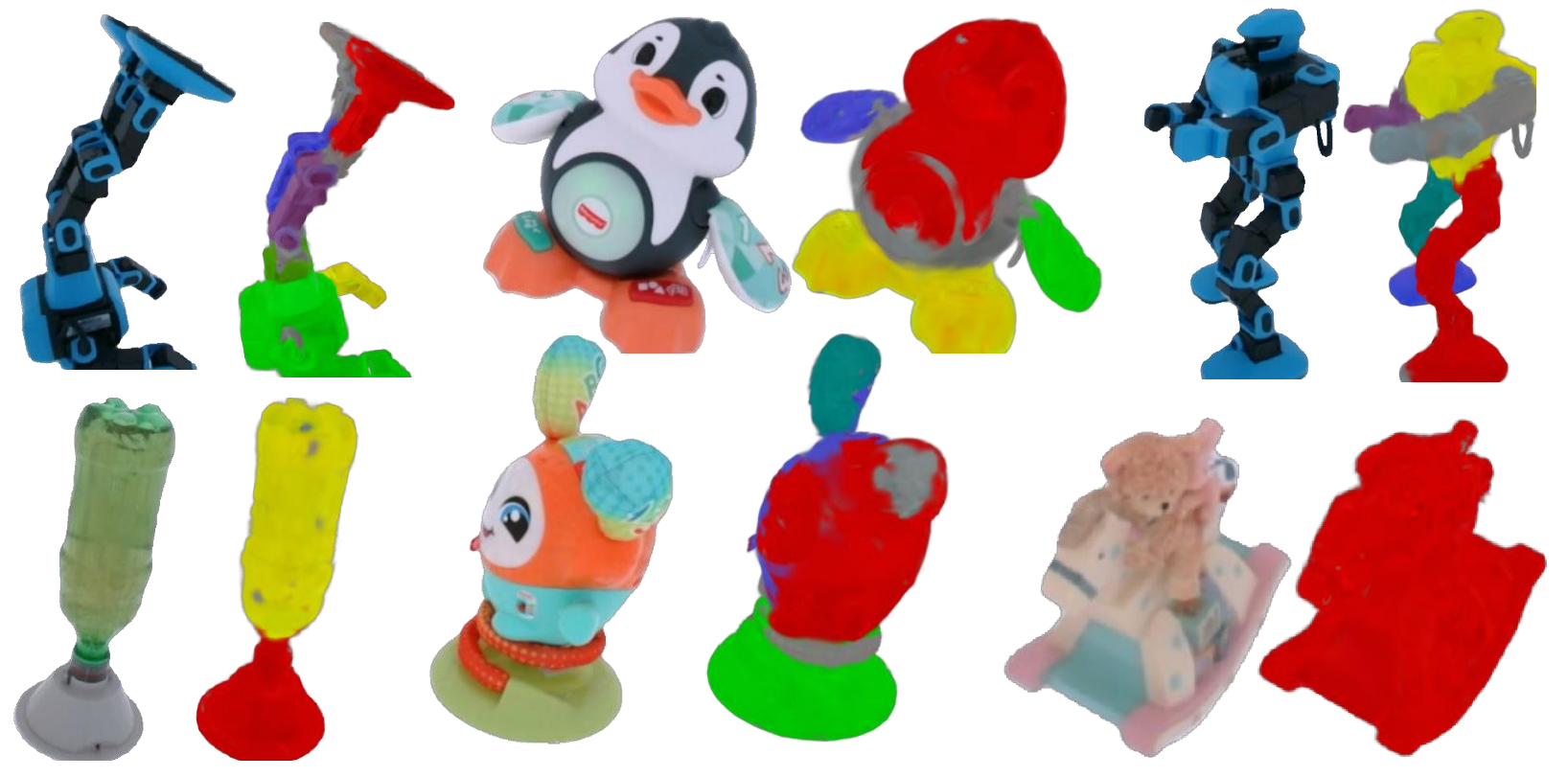}
    \caption{Rigid group visualization on Diva360 dataset. The rigid part of each object is colored.}
    \label{fig:diva_rigid_result}
\end{figure}

\begin{figure}
    \centering
    \includegraphics[width=\linewidth]{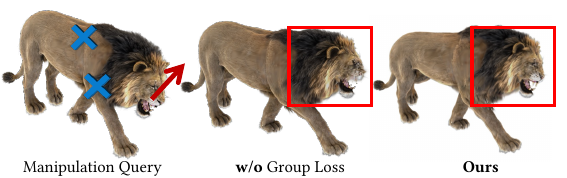}
    \caption{Additional manipulation example on DFA dataset.}
    \label{fig:another_manipulation_result}
\end{figure}

\section{Additional Experiment}\label{sec:supp_exp_set}

\subsection{Comparison of Optimization Time}

\begin{table}[h!]
\centering
\caption{\modified{Comparison of optimization time on Diva360 dataset.}}
\label{tab:method_time}
\resizebox{0.55\linewidth}{!}{
\textcolor{black}{
\begin{tabular}{l|r}
\toprule
\textbf{Method} & \textbf{Time} \\
\midrule
4DGS~\cite{wu20244dgs} & 96 min \\
SC-GS~\cite{huang2024scgs} & 98 min \\
GESI~\cite{luo2024gesi} & 10 min \\
3DGS~\cite{kerbl20233dgs} & \textbf{44 s} \\
Ours (w/ GT campose) & \underline{82 s} \\
\midrule
Ours (w/o GT campose) & 5 min \\
\bottomrule
\end{tabular}
}
}
\end{table}

\modified{
Table \ref{tab:method_time} compares computational cost with various baselines. We evaluate performance of our method under two conditions: with and without GT camera poses. It is important to note that other methods are all tested with GT poses provided. Under this comparable condition, our method takes an average of 82 seconds, making it faster than every other method except for 3DGS. In contrast, 4DGS and SC-GS require more time because they train a Multi-Layer Perceptron (MLP) for deformation, while GESI is time-consuming as it repeatedly performs expensive image matching. When GT poses are not provided, our method takes longer because it must perform image matching on numerous image pairs for camera pose selection.
}

\subsection{Robustness analysis}
\
\begin{table}[h!]
\caption{\modified{Robustness analysis of our method on the Diva360 dataset.}}
\centering
\label{tab:robustness}
\resizebox{0.95\linewidth}{!}{
\modified{
\resizebox{\columnwidth}{!}{
\begin{tabular}{lccc}
\toprule
Method Variant & PSNR$\uparrow$ & SSIM$\uparrow$ & LPIPS$\downarrow$ \\
\midrule
ours w/ 1k noise to Gaussians & 25.38 & 0.942 & 0.056 \\
ours w/ half training views & 25.50 & 0.947 & 0.056 \\
full pipeline (ours) & \textbf{26.84} & \textbf{0.955} & \textbf{0.050} \\
\bottomrule
\end{tabular}
}
}
}
\end{table}

\modified{
To verify robustness of our method, we also evaluated under two degraded 3DGS initialization.
}
\modified{
\paragraph{Noise to Gaussians.}
To verify robustness of our method, we introduced 1,000 random noise Gaussians to the initial 3D Gaussian model. The noise was sampled from a normal distribution and scaled based on mean and std of the original Gaussian parameters. As shown in Table \ref{tab:robustness}, our method exhibits only a minimal PSNR decrease of 1.46 dB, confirming its effectiveness even with noisy perturbation.
}

\modified{
\paragraph{Half Training Views.}
To test robustness under limited input, we initialized the Gaussian model with only half of the training views. As shown in Table \ref{tab:robustness}, this resulted in a minimal PSNR drop of just 1.34 dB, a performance that remains significantly higher than the previous state-of-the-art method, GESI~\cite{luo2024gesi}. This confirms strong resilience our method on various initial model quality.
}

\subsection{Multiple Time Selection}
\begin{table}[h!]

\caption{\modified{Evaluation Result of 5 randomly selected target timestep on Diva360 dataset.}}
\modified{
\centering
\label{tab:evaluation}
\resizebox{0.68\linewidth}{!}{
\begin{tabular}{lccc}
\toprule
Method & PSNR$\uparrow$ & SSIM$\uparrow$ & LPIPS$\downarrow$ \\
\midrule
3DGS~\cite{kerbl20233dgs}      & 21.30 & 0.898 & 0.098 \\
GESI~\cite{luo2024gesi}      & 22.54 & 0.913 & 0.085 \\
GESI($\mu,q$)~\cite{luo2024gesi} & 22.63 & 0.919 & 0.080 \\
Ours      & \textbf{25.17} & \textbf{0.939} & \textbf{0.063} \\
\bottomrule
\end{tabular}
}
}
\end{table}

\modified{
To demonstrate that our method is not limited to manually selected targets but also generalizes well to arbitrary target frames, we conducted experiments on 5 randomly chosen timesteps for target image seleciton.  As shown in Table~\ref{tab:evaluation}, our method outperforms the baselines by a margin of 2.5 dB in PSNR, highlighting its superior capability in deforming toward diverse target frames.
}

\section{Implementation Details and Hyperparameters}

\modified{
We implement our proposed method using the \texttt{gsplat}~\cite{ye2025gsplat} library. All experiments are carried out on an NVIDIA RTX 4090 GPU to ensure consistent computational performance. For the image matching stage, we remove spurious correspondences that do not belong to the target object. Specifically, we leverage the object masks provided by the datasets to filter out any matches falling outside the annotated object regions, thereby improving the reliability of the matching process.
}

\begin{table}[h!]
\centering
\caption{\modified{Hyperparameters for DFA and Diva360 datasets.}}
\label{tab:hyperparams}
\modified{
\resizebox{0.55\linewidth}{!}{
\begin{tabular}{l|c|c}
\toprule
 Parameter  & DFA & Diva360 \\
\midrule
$lr_q$ & 0.03 & 0.05 \\
$lr_t$ & 0.003 & 0.01 \\
$\tau_{\text{high}}$ & 0.01 & 0.01 \\
$\tau_{\text{low}}$ & 0.01 & 0.01 \\
$r_{\text{refinement}}$ & 0.05 & 0.01 \\
$k_{\text{anchor}}$ & 9 & 10 \\
$s_{\text{voxel}}$ & 0.02 & 0.06 \\
\bottomrule
\end{tabular}
}
}
\end{table}

\modified{
The hyperparameters employed in our experiments are listed in detail in Table~\ref{tab:hyperparams}.
$lr_q$ and $lr_t$ denote the learning rates applied to the rotation and translation of anchors, respectively.  
$\tau_{\text{high}}$ and $\tau_{\text{low}}$ are the refinement threshold used in the rigid part refinement stage.  
$r_{\text{refinement}}$ is the radius used in the BallQuery operation during refinement, while $k_{\text{anchor}}$ denotes the number of neighboring anchors in Eq.~2. $s_\text{voxel}$ is size of voxel for anchor initialization. 
}

\modified{
Hyperparameters are tuned using Bayesian optimization provided by \texttt{wandb}~\cite{biewald2020experiment}. For further minor detail of hyperparameters, please refer to our released code.
}




\end{document}